\documentclass[aip,graphicx]{revtex4-1}
\usepackage[cp1251]{inputenc}
\usepackage[T2A]{fontenc}
\usepackage[english]{babel}
\usepackage{amssymb,latexsym,amsmath,amscd}
\usepackage{graphicx,color,framed}
\usepackage{xcolor}
\usepackage{siunitx} 
\graphicspath{{figures/}}
\newcommand{\angstrom}{\mbox{\normalfont\AA}}
\begin{document}
\selectlanguage{english}
\title{Theory of ionic liquids with polarizable ions on a charged electrode}
\author{\firstname{Yury A.} \surname{Budkov}}
\email[]{ybudkov@hse.ru}
\affiliation{School of Applied Mathematics, HSE University, Tallinskaya st. 34, 123458 Moscow, Russia}
\affiliation{G.A. Krestov Institute of Solution Chemistry of the Russian Academy of Sciences, 153045, Akademicheskaya st. 1, Ivanovo, Russia}
\author{\firstname{Semen V.} \surname{Zavarzin}}
\affiliation{School of Applied Mathematics, HSE University, Tallinskaya st. 34, 123458 Moscow, Russia}
\author{Andrei L. Kolesnikov}
\affiliation{Institut f\"ur Nichtklassische Chemie e.V., Permoserstr. 15, 04318 Leipzig, Germany}
\begin{abstract}
We formulate a general mean-field theory of a flat electric double layer in ionic liquids and electrolyte solutions with ions possessing static polarizability and a permanent dipole moment on a charged electrode. We establish a new analytical expression for electric double layer differential capacitance determining it as an absolute value of the ratio of the local ionic charge density to the local electric field on an electrode surface. We demonstrate that this expression generalizes the analytical expressions previously reported by Kornyshev and Maggs and Podgornik. Using the obtained analytical expression, we explore new features of the differential capacitance behavior with an increase in the static polarizability and permanent dipole moment of cations. We relate these features to the behavior of ionic concentrations on the electrode. In particular, we elucidate the role of the competition between the dielectrophoretic attraction and Coulomb repulsion forces acting on polarizable or polar cations in the electric double layer in the behavior of the differential capacitance. The developed theoretical model and obtained theoretical findings could be relevant for different electrochemical applications, e.g. batteries, supercapacitors, catalysis, electrodeposition, {\sl etc}.
\end{abstract}
\maketitle
\section{Introduction}
Understanding the behavior of differential capacitance of an electric double layer (EDL) occurring in room temperature ionic liquids (RTILs) on the charged conducting electrodes and its dependence on applied voltage is one of the main problems of electrochemistry of RTILs \cite{fedorov2014ionic}. A lot of theoretical models of RTILs on electrified interfaces have been so far proposed: from the seminal phenomenological mean-field theory \cite{kornyshev2007double} to sophisticated statistical models based on the self-consistent field theory \cite{maggs2016general,budkov2018theory,caetano2017differential,may2019differential,cruz2019effect} and the density functional theory \cite{jiang2011density,wu2011classical,forsman2011classical,frydel2012close} aiming at studying the differential capacitance behavior.

A lot of RTILs consist of molecular organic cations carrying not only a certain electric charge, but also a sufficiently large permanent dipole moment \cite{izgorodina2009components,schroder2006simulation,prado2006molecular} and having high static electronic polarizability \cite{schroder2010simulating,mcdaniel2018influence} (in what follows referred to as static polarizability) and inorganic anions with relatively small static polarizability. As it was shown in numerous papers \cite{schroder2006simulation,schroder2010simulating,mcdaniel2018influence}, polarizability and dipole moment of cations strongly affect the local structure of bulk RTILs.

Based on the general considerations, apart from the influence on the bulk local structure, the presence of the permanent dipole moment or the static polarizability on the organic cations should affect the differential capacitance of EDL on ionic liquid-charged electrode interfaces. Indeed, polarizable or polar organic cations of an ionic liquid being in an inhomogeneous electric field near a charged electrode are exposed to the action of a dielectrophoretic force \cite{jones1979dielectrophoretic,budkov2018theory} always directed towards the electrode surface, where the local electric field reaches a maximal value. Therefore, we can suppose that an additional attractive dielectrophoretic force acting on the cations should lead to a higher total ionic charge accumulated in the EDL and, thereby, to higher differential capacitance than what is predicted within the model without an explicit account of ion static polarizability \cite{kornyshev2007double,maggs2016general}. In order to understand how induced or permanent dipole moments of ions affect differential capacitance as a function of applied voltage, it is necessary to formulate an analytical self-consistent field theory of EDL in RTILs, whose ions possess static polarizability and permanent dipole moment.

To the best of our knowledge, up to now, the ion static polarizability effect on the differential capacitance in liquid electrolytes on charged electrodes has been discussed only in two papers \cite{lauw2009room,hatlo2012electric}. When modeling cations and anions as segmented dendrimers and calculating the effective dielectric constant from the local distribution of ions in order to take into account the excluded volume and the local dielectric screening effects, within the numerical self-consistent field theory, the authors obtained the resulting camel-shaped capacitance dependence on voltage and analyzed it in terms of the thickness of alternating layers and polarization of ions at electrified interfaces \cite{lauw2009room}. The authors of paper \cite{lauw2009room} included the excess ion polarizability effect in the Poisson-Boltzmann theory for aqueous electrolyte solutions on a charged electrode. Based on the proposed modified Poisson-Boltzmann theory, they showed that the decrease in differential capacitance with voltage, observed in metal electrodes above a threshold value of voltage (camel-shaped profile), can be understood in terms of double layer thickening due to ion-induced polarizability holes in water. The authors introduced a new length scale related to the excess ion polarizability of ions and showed that when it is comparable to the effective Debye length, ion static polarizability can significantly influence double layer properties. It is also important to discuss several papers \cite{frydel2011polarizable,buyukdagli2013microscopic,demery2012electrostatic,budkov2020two}, where the static polarizability or permanent dipole moment of ions were taken into account in the context of the theory of liquid electrolytes on charged interfaces. In paper \cite{frydel2011polarizable}, the author incorporated ion static polarizabilities into the Poisson-Boltzmann equation by modifying the effective dielectric constant and the Boltzmann distribution of ions and thus formulated the so-called polarizable Poisson-Boltzmann equation. Performing a fine analytical analysis based on the contact value theorem, the author concluded that the effect of ion static polarizability is relevant only to electrolytes dissolved in low-polar solvents and to ionic systems with a high ionic concentration like RTILs. However, the formulated polarizable Poisson-Boltzmann equation does not take into account the excluded volume interactions of ions and, strictly speaking, cannot be applied to description of EDLs in ionic liquids and concentrated electrolyte solutions on charged electrodes. In paper \cite{buyukdagli2013microscopic}, the authors introduced a field-theoretical model of a polar liquid composed of linear multipole solvent molecules and embedding polarizable ions modeled as Drude oscillators. The authors demonstrated that in contrast to the previous dipolar Poisson-Boltzmann formulations \cite{abrashkin2007dipolar,coalson1996statistical} treating solvent molecules as point-like dipoles, their model is able to qualitatively reproduce the nonlocal dielectric response behavior of polar liquids observed in molecular dynamics simulations and Atomic Force Microscopy experiments for an aqueous solvent on charged interfaces. The authors also showed within their field-theoretical model that ion static polarizability reverses the local ionic concentration trends predicted by the classical Poisson-Boltzmann theory, resulting in surface affinity of coions and exclusion of counterions. In paper \cite{demery2012electrostatic}, the authors carried out field-theoretical treatment of an inhomogeneous Coulomb fluid consisting of charged particles with static polarizability. The authors derived weak- and strong-coupling approximations and evaluated the partition function of a fluid confined between flat charged dielectric walls. They investigated the density profiles and the disjoining pressure for both approximations. A comparison with the case of nonpolarizable counterions showed that static polarizability of ions brings important differences to the counterion density distribution as well as the counterion mediated electrostatic disjoining pressure. In paper \cite{budkov2020two}, the authors developed a self-consistent field theory of a two-component electrolyte solution on a negatively charged electrode (cathode), predicting the efficient replacement of simple (alkali) cations with dipolar (organic) ones within the EDL. For the typical values of the molecular dipole moment, the replacement manifests itself at reasonable surface charge densities of the electrode.

In this paper, we will propose a general mean-field theory of a flat EDL in RTILs and rather dilute electrolyte solutions with ions possessing static polarizabilities and a permanent dipole moments on a charged electrode. We will explore new features of the differential capacitance behavior by increasing the static polarizability and permanent dipole moment of cations. Then, we will discuss the role of the competition between dielectrophoretic attraction and Coulomb repulsion acting on polarizable or polar cations within the EDL in the behavior of the differential capacitance as a function of an applied voltage.

\section{General theory}
Let us consider an ionic liquid comprised of ions with charges $\pm q$ in the thermodynamic equilibrium state at a certain temperature $T$. We assume that an ion of each kind possesses, in the general case, a permanent dipole moment, $p_{\pm}$, and static polarizability, $\alpha_{\pm}$, where the $\pm$ symbols refer to the cation and anion, respectively. A flat charged electrode with a maintained potential drop is immersed in an ionic liquid, so that an electric double layer (EDL) is formed in its vicinity. Placing the origin of the $z$-axis on the electrode surface, we can write the grand thermodynamic potential (GTP) of the ionic liquid per unit electrode area as follows
\begin{equation}
\label{omega}
\Omega/\mathcal{A}=\int\limits_{0}^{\infty}\left(-\frac{\varepsilon\varepsilon_{0} \mathcal{E}^2}{2}+\rho\psi-c_{+}\Psi_{+}-c_{-}\Psi_{-}+f-\mu_{+}c_{+}-\mu_{-}c_{-}\right)dz,
\end{equation}
where $\mathcal{A}$ is the total area of the electrode, $\psi(z)$ is the local electrostatic potential and $\mathcal{E}(z)=-\psi^{\prime}(z)$ is the local electric field, $\rho(z)=q\left(c_{+}(z)-c_{-}(z)\right)$ is the local charge density of the ions, $\varepsilon$ is the reference dielectric permittivity of the ionic liquid which is an adjustable parameter, not dependent on the polarizabilities or dipole moments of ionic species and determined by the unaccounted effects, such as cluster formation\cite{avni2020charge,zhang2020enforced,feng2019free}, $\varepsilon_0$ is the permittivity of the empty space, $f=f(T,c_{+},c_{-})$ is the intrinsic free energy density of the ionic liquid which in the local density approximation is simply a function of the local ionic concentrations, $c_{\pm}$, and the temperature, $T$, $\mu_{\pm}$ are the bulk chemical potentials of the species. The first and second terms in the integrand of (\ref{omega}) describe the electrostatic free energy of the ionic liquid in the mean-field approximation. The third and fourth terms determine the free energy density of particles with induced and freely rotating permanent dipoles in the electric field, $\mathcal{E}$, with the following auxiliary functions \cite{budkov2018theory,budkov2016theory}
\begin{equation}
\Psi_{\pm}(z)=\frac{\alpha_{\pm}\mathcal{E}^2(z)}{2}+k_{B}T\ln\left(\frac{\sinh{\beta p_{\pm}\mathcal{E}(z)}}{\beta p_{\pm}\mathcal{E}(z)}\right).
\end{equation}
In this study, we would like to focus only on the effects of electronic and orientation polarizabilities of ions on the differential capacitance of EDL and, for simplicity, we do not take into account the short-range specific interactions \cite{budkov2018theory,goodwin2017mean,downing2018differential,zhao2012influence} and structural interactions \cite{blossey2017structural} between the ionic species and their specific adsorption onto the electrode leading to the Stern layer formation \cite{budkov2018theory,uematsu2018effects,podgornik2018general}. Accounting for the specific interactions between ions and their specific adsorption onto the electrode allows to arrive at capacitance values that are comparable to the experimental values \cite{budkov2018theory}. Moreover, we do not explicitly consider the formation of ionic pairs \cite{zhang2020enforced} contributing to the dielectric permittivity and quadrupolar ionic clusters resulting in the electrostatic potential oscillations at long distances from the electrode \cite{avni2020charge} or point-like test charge \cite{budkov2020statistical,budkov2019statistical}. Finally, we do not take into account the electrostatic correlations leading to the nonlocal terms in the electrostatic free energy functional \cite{bazant2011double,misra2019theory}.

Using the Legendre transformation \cite{budkov2016theory,budkov2018theory,budkov2020two,maggs2016general} 
\begin{equation}
P(T,\mu_{+},\mu_{-})=\mu_{+}c_{+}+\mu_{-}c_{-}-f(T,c_{+},c_{-}),~\mu_{\pm}=\frac{\partial f}{\partial c_{\pm}},   
\end{equation}
we rewrite the GTP as the following functional of the electrostatic potential 
\begin{equation}
\label{therm_pot}
\Omega[\psi]/\mathcal{A}=-\int\limits_{0}^{\infty}dz\left(\frac{\varepsilon\varepsilon_0 \mathcal{E}^2(z)}{2}+P(T,\bar{\mu}_{+}(z),\bar{\mu}_{-}(z))\right), \end{equation}
where
\begin{equation}
\label{int_chem_pot}
\bar{\mu}_{\pm}=\mu_{\pm}\mp\ q\psi+\frac{\alpha_{\pm}\mathcal{E}^2}{2}+k_{B}T\ln\left(\frac{\sinh{\beta p_{\pm}\mathcal{E}}}{\beta p_{\pm}\mathcal{E}}\right)
\end{equation}
are the intrinsic chemical potentials of the ionic species, $\beta=(k_{B}T)^{-1}$, $k_{B}$ is the Boltzmann constant, $P(T,\bar{\mu}_{+},\bar{\mu}_{-})$ is the local pressure of ions as a function of intrinsic chemical potentials of species. We would like to note that eqs. (\ref{int_chem_pot}) determine the condition of the thermodynamic equilibrium of an inhomogeneous system \cite{landau2013course}.

In order to obtain the potential profile, $\psi(z)$, it is necessary to solve the self-consistent field equation
\begin{equation}
\label{scfe}
\frac{d}{dz}\left(\epsilon(z)\psi^{\prime}(z)\right)=-q\left(\bar{c}_{+}(z)-\bar{c}_{-}(z)\right),
\end{equation}
which can be obtained by varying the thermodynamic potential (\ref{therm_pot}) with respect to the electrostatic potential \cite{abrashkin2007dipolar,budkov2015modified}, where 
\begin{equation}
\epsilon(z)=\varepsilon\varepsilon_0+\sum\limits_{i=\pm}^{}\left(\alpha_{i}+\frac{p_{i}^2}{k_{B}T}\frac{L(\beta p_{i}\mathcal{E}(z))}{\beta p_{i}\mathcal{E}(z)}\right)\bar{c}_{i}(z),
\end{equation}
is the effective local dielectric permittivity of the fluid taking into account the effect of electronic and orientational polarizabilities of the ions, $L(x)=\coth{x}-1/x$ is the well-known Langevin function, and
\begin{equation}
\label{local_conc}
\bar{c}_{\pm}(z)=\frac{\partial P}{\partial{\bar{\mu}_{\pm}}(z)}
\end{equation}
are the equilibrium average ionic concentrations which depend on the coordinate via the intrinsic chemical potentials, $\bar{\mu}_{\pm}(z)$. The self-consistent field equation (\ref{scfe}) is the well-known Poisson equation for the potential of the self-consistent electrostatic field. For the bulk solution, where $c_{\pm}=c$, we have the following mean-field expression for the bulk dielectric permittivity
\begin{equation}
\label{bulk_perm}
\epsilon=\varepsilon\varepsilon_0+\sum\limits_{i=\pm}^{}\left(\alpha_{i}+\frac{p_{i}^2}{3k_{B}T}\right)c.
\end{equation}
We would also like to note that the dipole moments, $p_{\pm}$, the polarizabilities, $\alpha_{\pm}$, and the reference dielectric permittivity, $\varepsilon$, in eq. (\ref{bulk_perm}) can be considered as effective adjustable parameters to fit the experimentally measured bulk dielectric permittivity of a certain ionic liquid (see also \cite{abrashkin2007dipolar}).

We would like to note that for the case of the ideal gas reference model with the equation of state $P(T,\mu_{+},\mu_{-})=k_{B}T\left(\lambda_{+}^{-3}e^{\beta\mu_{+}}+\lambda_{-}^{-3}e^{\beta\mu_{-}}\right)$ with the de Broglie thermal wavelengths of the ions, $\lambda_{\pm}$, the general self-consistent field equation (\ref{scfe}) transforms into the polarizable Poisson-Boltzmann equation \cite{frydel2011polarizable} mentioned in the Introduction.

The boundary conditions for the self-consistent field equation are
\begin{equation}
\psi(0)=\psi_{0},~\mathcal{E}(\infty)=0,
\end{equation}
where $\psi_0$ is the potential drop on the electrode. The first integral of the self-consistent field equation \cite{budkov2015modified,budkov2016theory,abrashkin2007dipolar,misra2019theory}, which describes the mechanical stability of the fluid, is
\begin{equation}
\label{equil}
-\frac{\varepsilon\varepsilon_0 \mathcal{E}^2}{2}-\sum\limits_{i=\pm}\mathcal{E}\left(\alpha_{i}\mathcal{E}+p_{i}L(\beta p_{i}\mathcal{E})\right)\bar{c}_{i}+P(T,\bar{\mu}_{+},\bar{\mu}_{-})=P_{0},
\end{equation}
where $P_{0}=P(T,\mu_{+},\mu_{-})$ is the pressure of the ions in the bulk. We would like to note that eq. (\ref{equil}) can be obtained from the well-known contact value theorem \cite{demery2012electrostatic,frydel2011polarizable} as well. Further, to calculate the differential capacitance of the EDL, $C={d \sigma}/d{\psi_0}$, which is the main subject of this paper, it is necessary to calculate the electrode surface charge density using the well-known expression \cite{landau2013electrodynamics} $\sigma=\epsilon_s \mathcal{E}_s$ with the local dielectric permittivity
\begin{equation}
\label{loc_eps}
\epsilon(0)=\epsilon_s=\varepsilon\varepsilon_0+\sum\limits_{i=\pm}^{}\left(\alpha_{i}+\frac{p_{i}^2}{k_{B}T}\frac{L(\beta p_{i}\mathcal{E}_s)}{\beta p_{i}\mathcal{E}_s}\right){c}_{s,i},
\end{equation}
whereas the local electric field, $\mathcal{E}_s=\mathcal{E}(0)$, on the electrode can be determined from the mechanical equilibrium condition (\ref{equil}) at $z=0$, i.e.
\begin{equation}
\label{loc_eq}
-\frac{\varepsilon\varepsilon_0 \mathcal{E}_s^2}{2}-\sum\limits_{i=\pm}\mathcal{E}_s\left(\alpha_{i}\mathcal{E}_s+p_{i}L(\beta p_{i}\mathcal{E}_s)\right){c}_{s,i}+P_s=P_{0},
\end{equation}
where ${c}_{s\pm}=\bar{c}_{\pm}(0)={c}_{s\pm}(T,{\mu}_{s+},{\mu}_{s-})$ and $P_{s}=P(T,{\mu}_{s+},{\mu}_{s-})$ are the local ionic concentrations and pressure of ions on the electrode, respectively, which are the functions of the local intrinsic chemical potentials of the ions
\begin{equation}
\label{loc_int_chem_pot}
{\mu}_{s\pm}=\mu_{\pm}\mp q\psi_0+\frac{\alpha_{\pm}\mathcal{E}_s^2}{2}+k_{B}T\ln\left(\frac{\sinh{\beta p_{\pm}\mathcal{E}_s}}{\beta p_{\pm}\mathcal{E}_s}\right).
\end{equation}
Then, calculating the derivative $d{\sigma}/d{\psi}_0=(\partial{\sigma}/\partial{\psi}_0)_{\mathcal{E}_s}+(\partial{\sigma}/\partial{\mathcal{E}_s})_{\psi_0}(d{\mathcal{E}_s}/d{\psi}_0)$ with the help of eqs. (\ref{loc_eps}) and (\ref{loc_eq}), taking into account that the local ionic concentrations, $c_{s\pm}$, depend on the surface potential, $\psi_0$, and local electric field, $\mathcal{E}_{s}$, via the local intrinsic chemical potentials (\ref{loc_int_chem_pot}), after some cumbersome algebraic calculations, we arrive at the following simple analytical expression for the differential capacitance
\begin{equation}
\label{cap}
C=\frac{q\left({c}_{s-}-{c}_{s+}\right)}{\mathcal{E}_s}=-\frac{\rho_{s}}{{\mathcal{E}_{s}}},
\end{equation}
where the local electric field on the electrode $\mathcal{E}_s=\mathcal{E}_s(\psi_0)$ can be calculated from the numerical solution of eq. (\ref{loc_eq}) and $\rho_{s}(\psi_0)=q\left({c}_{s+}(\psi_0)-{c}_{s-}(\psi_0)\right)$ is the local ionic charge density on the electrode. Eq. (\ref{cap}) as the main result of this paper can be used to estimate the differential capacitance within the mean-field approximation for an arbitrary equation of state, $P=P(T,\mu_{+},\mu_{-})$, of the reference ionic liquid model taking into account the permanent dipole moments and the static polarizabilities of ions. We would like to note that eq. (\ref{cap}) can be obtained using general physical arguments. Namely, 
\begin{equation}
C=\frac{d\sigma}{d\psi_0}=\frac{dD(0)}{d\psi(0)}=\frac{\left(\frac{dD}{dz}\right)}{\left(\frac{d\psi}{dz}\right)}\biggr\rvert_{z=0}=-\frac{\rho_{s}}{\mathcal{E}_s},
\end{equation}
where we have used the Gauss law, $dD/dz=q(c_{+}(z)-c_{-}(z))$, and the definition of the local electric field, $\mathcal{E}(z)=-\psi^{\prime}(z)$.

Note that for the case, when $p_{\pm}=\alpha_{\pm}=0$, eq. (\ref{cap}) transforms into the expression obtained for the first time by Maggs and Podgornik \cite{maggs2016general}
\begin{equation}
C=\pm\frac{q\varepsilon\varepsilon_0\left({c}_{s-}-{c}_{s+}\right)}{\sqrt{2\varepsilon\varepsilon_0\left(P_s-P_0\right)}},
\end{equation}
with $\pm$ signs corresponding to the sign of the electrode charge. 
We would also like to note that two of us earlier, in paper \cite{budkov2015modified}, applied a different approach to obtain an analytical expression, similar to eq. (14), for the differential capacitance of a dilute electrolyte solution with a small cosolvent additive with polarizable or polar molecules, where the ions and cosolvent molecules were modeled as point-like particles.

\section{Symmetric lattice gas reference system}
Before the numerical calculations, we would like to describe the reference ionic liquid model with a certain equation of state $P=P(T,\mu_{+},\mu_{-})$. In the present study, we do not consider the effects of asymmetric steric interactions on the differential capacitance which were studied recently in paper \cite{maggs2016general} within the Carnahan-Starling approximation for a hard spheres mixture and a two-component asymmetric lattice gas model. In this study, we use a two-component symmetric lattice gas model with the following dependence of pressure on the chemical potentials of ionic species, as the reference ionic liquid model \cite{hill1986introduction,maggs2016general,budkov2016theory}
\begin{equation}
P(T,\mu_{+},\mu_{-})=\frac{k_{B}T}{v}\ln\left(1+e^{\beta\mu_{+}}+e^{\beta\mu_{-}}\right),    
\end{equation}
where $v$ is the volume of the lattice gas elementary cell. Calculating the derivative in (\ref{local_conc}), and using the expression for the bulk chemical potentials $\mu_{\pm}=k_{B}T\ln({cv}/(1-2cv))$, where $c$ is the introduced above bulk concentration of the ions, we arrive at the following expressions for the average ionic concentrations
 \begin{equation}
\bar{c}_{\pm}=c\frac{e^{\beta\left(\frac{\alpha_{\pm}\mathcal{E}^2}{2}\mp q\psi\right)}\frac{\sinh{(\beta p_{\pm}\mathcal{E}})}{\beta p_{\pm}\mathcal{E}}}{1+cv\left(e^{\beta\left(\frac{\alpha_{+}\mathcal{E}^2}{2}- q\psi\right)}\frac{\sinh{(\beta p_{+}\mathcal{E}})}{\beta p_{+}\mathcal{E}}+e^{\beta\left(\frac{\alpha_{-}\mathcal{E}^2}{2}+q\psi\right)}\frac{\sinh{(\beta p_{-}\mathcal{E}})}{\beta p_{-}\mathcal{E}}-2\right)}.
\end{equation}
Thus, the ionic concentrations on the electrode take the following form
\begin{equation}
\label{loc_conc}
c_{s\pm}=c\frac{e^{\beta\left(\frac{\alpha_{\pm}\mathcal{E}_{s}^2}{2}\mp q\psi_0\right)}\frac{\sinh{(\beta p_{\pm}\mathcal{E}_s})}{\beta p_{\pm}\mathcal{E}_s}}{1+cv\left(e^{\beta\left(\frac{\alpha_{+}\mathcal{E}_s^2}{2}- q\psi_0\right)}\frac{\sinh{(\beta p_{+}\mathcal{E}_s})}{\beta p_{+}\mathcal{E}_s}+e^{\beta\left(\frac{\alpha_{-}\mathcal{E}_s^2}{2}+q\psi_0\right)}\frac{\sinh{(\beta p_{-}\mathcal{E}_s})}{\beta p_{-}\mathcal{E}_s}-2\right)}.
\end{equation}
Eqs. (\ref{loc_eq}), (\ref{cap}), and (\ref{loc_conc}) form a closed system of coupled equations allowing us to calculate the differential capacitance as a potential drop function. Note that for the case of nonpolar and nonpolarizable ions ($\alpha_{\pm}=p_{\pm}=0$), eq. (\ref{cap}) with the help of eq. (\ref{loc_eq}) gives Kornyshev's well-known expression \cite{kornyshev2007double}
\begin{equation}
C=\frac{\left({2\beta\varepsilon\varepsilon_0c^2vq^2}\right)^{1/2}|\sinh{\beta q\psi_0}|}{\left(1+4cv\sinh^2\left(\frac{\beta q\psi_0}{2}\right)\right)\sqrt{\ln\left(1+4cv\sinh^2\left(\frac{\beta q\psi_0}{2}\right)\right)}}.
\end{equation}

\section{Numerical results and discussions}
In what follows, we will focus on the physically relevant cases of an ionic liquid and a dilute electrolyte solution with cations possessing nonzero static polarizability or a permanent dipole moment. This situation might be realized for ionic liquids and their solutions with organic cations possessing large static polarizability and highly polar groups \cite{izgorodina2009components,schroder2006simulation,prado2006molecular,schroder2010simulating,mcdaniel2018influence}. Thus, in this case we can neglect, for simplicity, the static polarizability of inorganic anions. 
\subsection{Ionic liquid case}
\begin{figure}
\centering
\begin{tabular}{c c}
\includegraphics[width=0.5\linewidth]{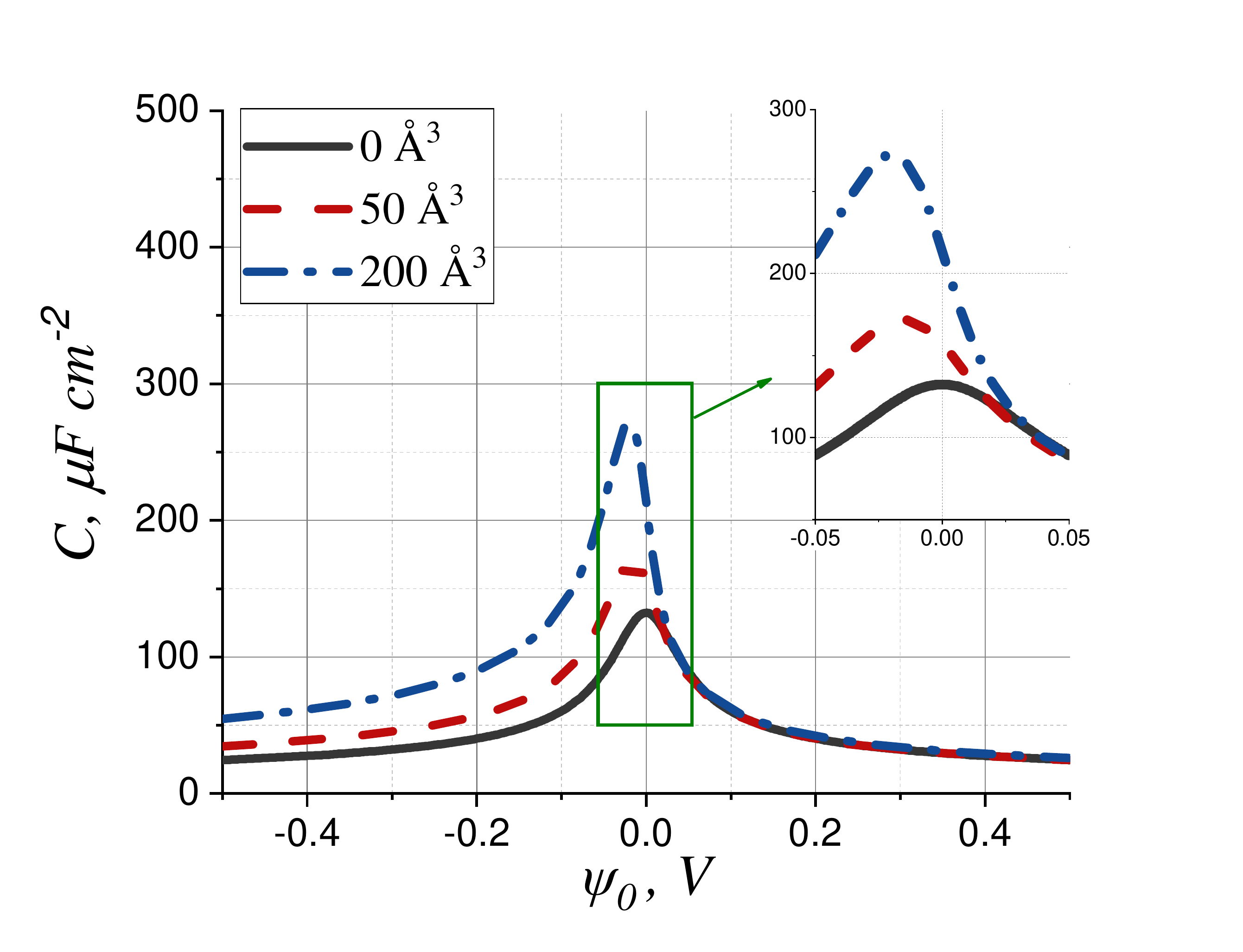}&\includegraphics[width=0.5\linewidth]{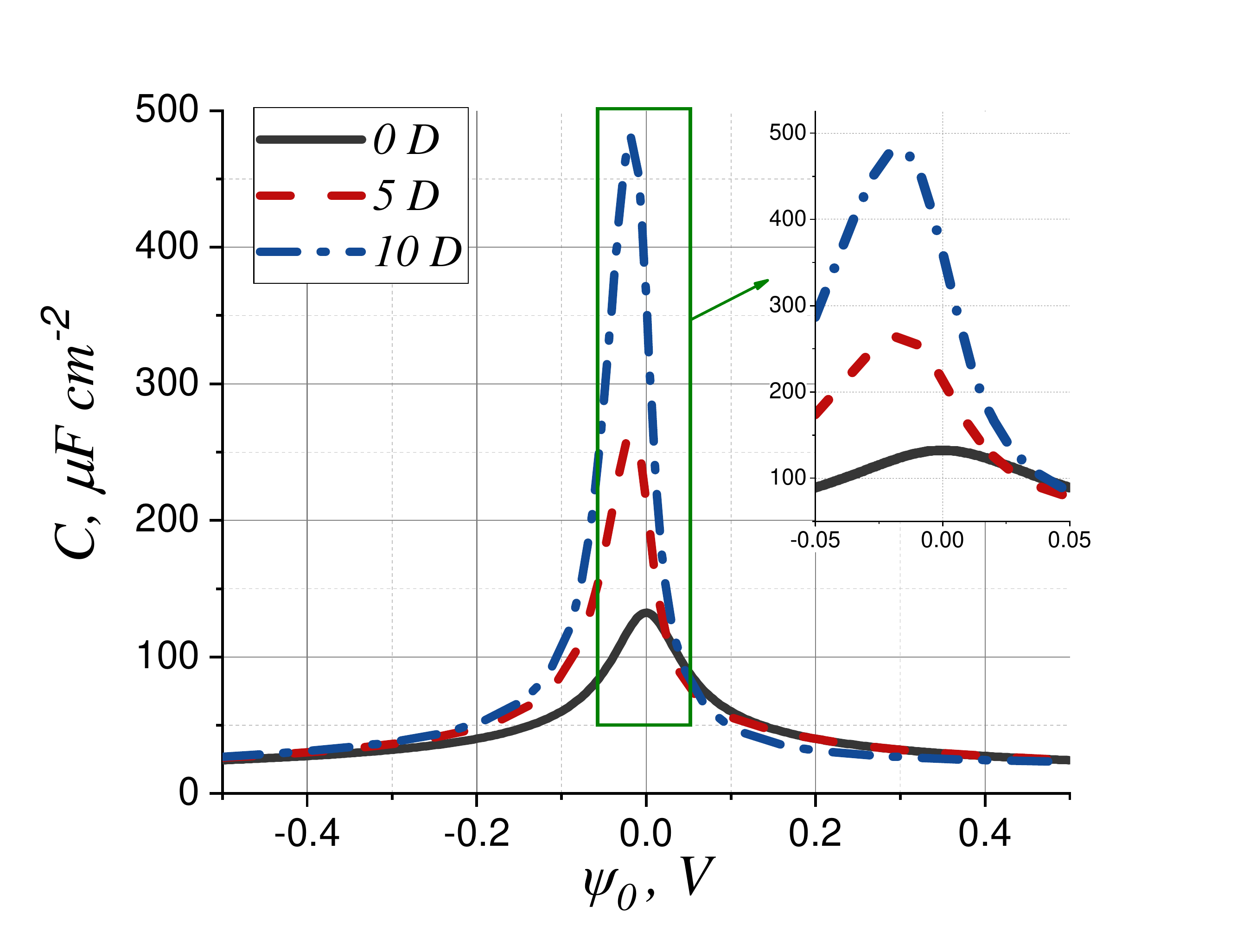} \\ [-0.3cm]
(a)&(b)\\
\end{tabular}
\caption{Differential capacitance of the EDL in an ionic liquid as a potential drop function at different values of static polarizability $\alpha_{+}$ (in $4\pi\varepsilon_0$ units) (a) and dipole moment $p_{+}$ (b). The data are shown for $\alpha_{-}=p_{+}=p_{-}=0$ (a), $\alpha_{-}=\alpha_{+}=p_{-}=0$ (b), $\varepsilon=5$, $v^{1/3}=5~{\angstrom}$, $cv = 0.4$, $T=298~K$.}
	\label{fig:CapIL}
\end{figure}
Fig. \ref{fig:CapIL}(a) shows the differential capacitance dependences on the potential drop, $\psi_0$, for an ionic liquid at the temperature $T=298~K$ and volume fraction $cv=0.4$ with $v^{1/3}=5~{\angstrom}$ at different values of static polarizability of cations, $\alpha_{+}$, expressed in $4\pi\varepsilon_0$ units. As is seen, at $\alpha_{+}=0$, the differential capacitance has a symmetric bell-shaped profile in accordance with Kornyshev's mean-field theory\cite{kornyshev2007double}. However, an increase in the electronic polarizability leads to higher differential capacitance at negative voltages and shifts the position of its maximum to more negative voltages (see Fig. 1a). As is seen from Fig.\ref{fig:CapIL}b, similar differential capacitance behavior at negative voltages is observed in case of polar cations -- the differential capacitance becomes higher with an increase in the permanent dipole moment of the cation, while the position of its maximum also shifts to more negative voltages. Note that the differential capacitance growth with an increase in the cation dipole moment at moderate negative potential drop values is more pronounced than that following an increase in the cation static polarizability. Nevertheless, at sufficiently high negative voltages, an increase in the dipole moment weakly influences on the differential capacitance value (see explanation below).
\begin{figure}
\begin{tabular}{c c}
\includegraphics[width=0.5\linewidth]{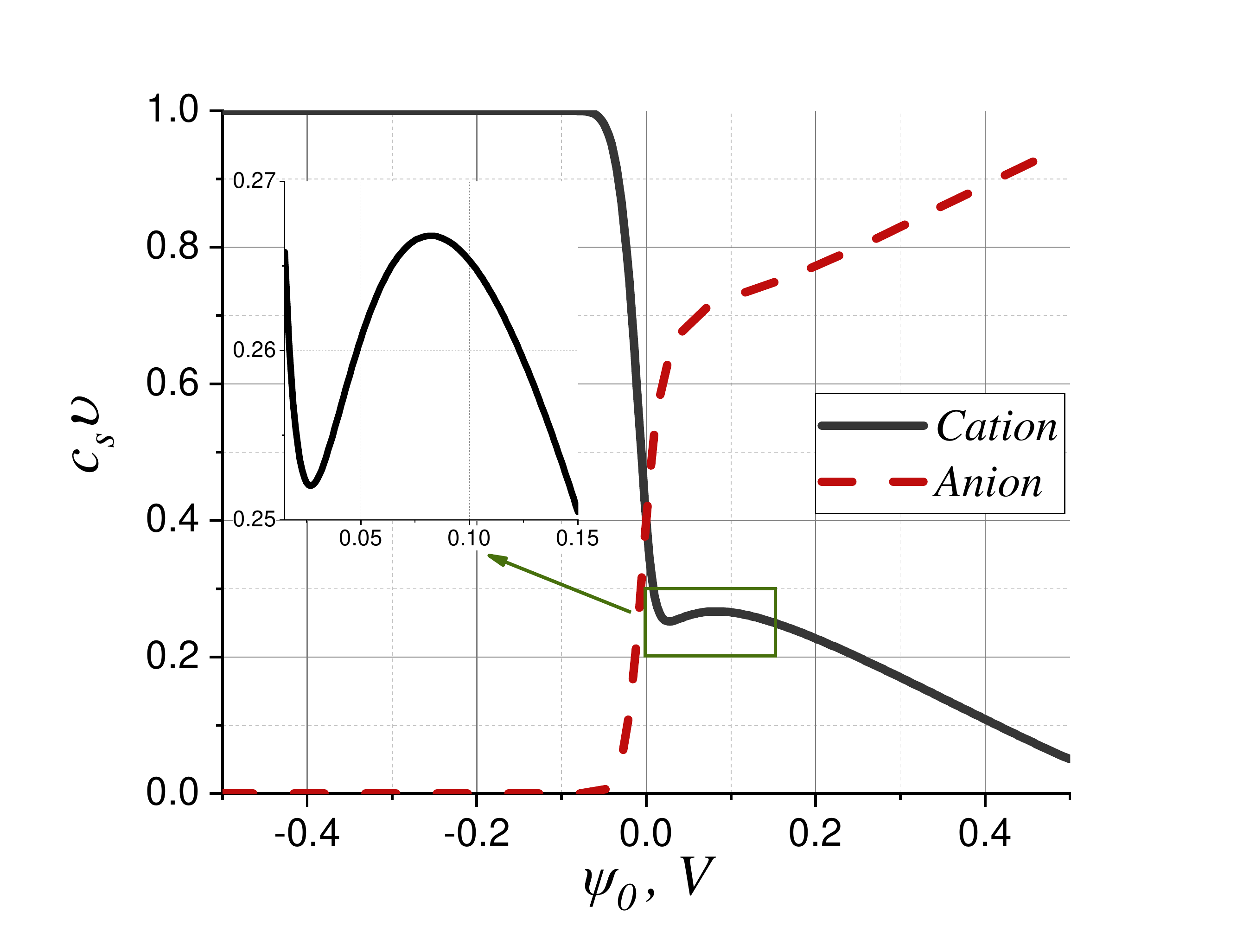}&\includegraphics[width=0.5\linewidth]{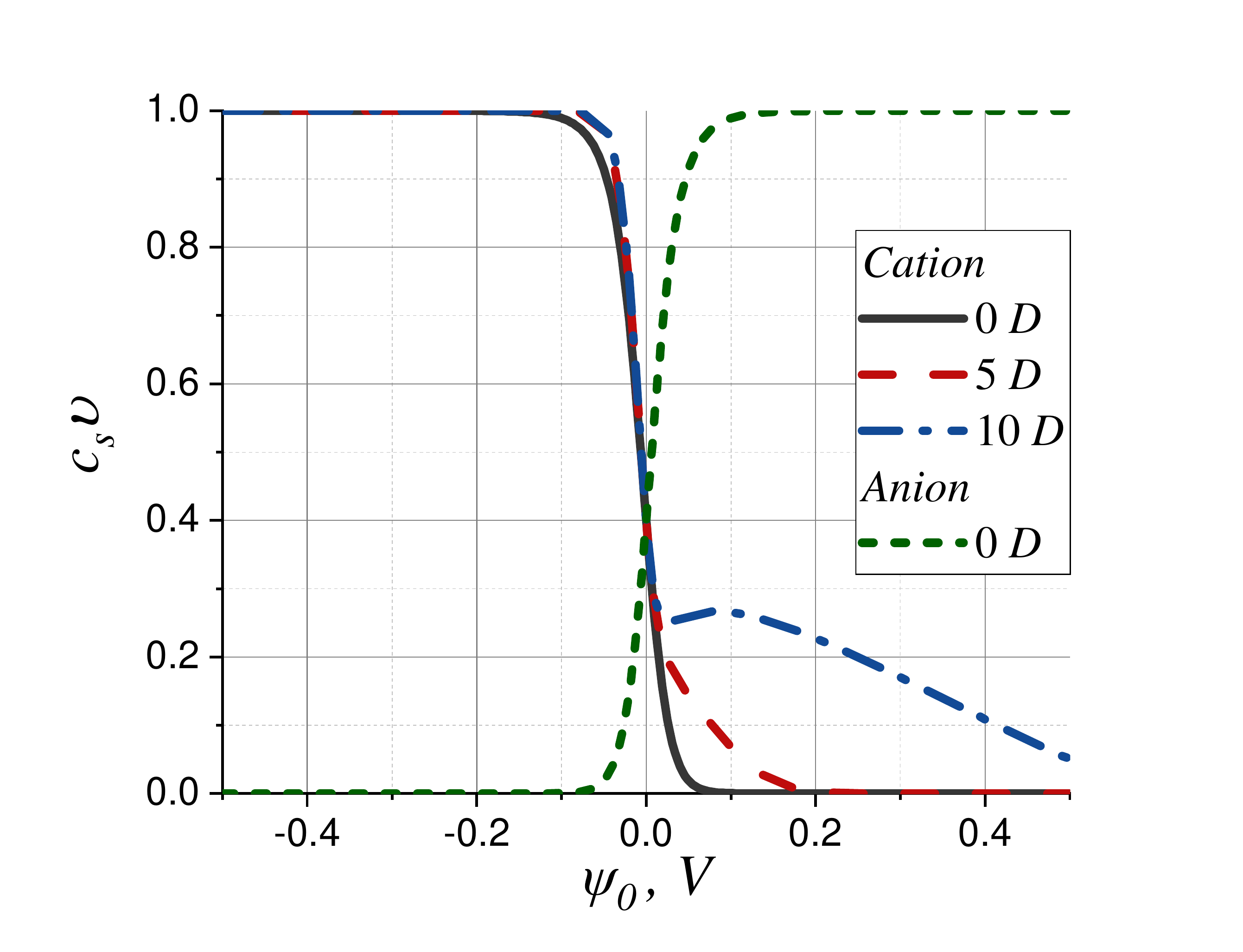} \\ [-0.3cm]
(a)&(b)\\
\end{tabular}
\caption{The volume fraction of cations and anions at dipole moment of cation $p_{+}=10~D$ (a) and the volume fraction of cations at different values of dipole moment $p_{+}$ (b) on the electrode. The data are shown for $\alpha_{-}=\alpha_{+}=p_{-}=0$, $\varepsilon=5$, $v^{1/3}=5~{\angstrom}$, $cv = 0.4$, $T=298~K$.}
\label{fig:ConcIL}
\end{figure}
The presence of static polarizability or a permanent dipole moment on the cations makes their attraction to the negatively charged electrode stronger than that of nonpolarizable anions to the positively charged electrode surface. The stronger attraction of the cations to the electrode is the result of an additional dielectrophoretic attractive force acting on the induced or permanent dipole placed in an inhomogeneous electric field \cite{jones1979dielectrophoretic,budkov2018theory,budkov2020two}. Fig. \ref{fig:ConcIL} shows the dependences of ionic concentrations on the electrode on the surface potential corresponding to the same physical parameters which were used for Fig. \ref{fig:CapIL}(a,b) in the case of a polar cation ($p_{+}\neq 0$, $\alpha_{+}=0$). As is seen, at a sufficiently large dipole moment ($p_{+}=10~D$), the competition between dielectrophoretic attraction and Coulomb repulsion acting on the polar cations makes the behavior of the local concentration of cations nonmonotonic. Such behavior can be explained as follows. At rather small voltages, the Coulomb repulsion of the cations from the electrode surface must be stronger than their dielectrophoretic attraction towards the electrode, which leads to a monotonical decrease in the local concentration of the cations. However, when the voltage exceeds a certain threshold value, corresponding to the local minimum of the cation concentration on the electrode, the dielectrophoretic force becomes larger than the Coulomb one. The latter increases in the local concentration of the cations with voltage. Nevertheless, at a sufficiently large voltage value, after a certain threshold value, the Coulomb repulsion of the cations from the electrode surface becomes stronger than their dielectrophoretic attraction again, and the concentration of the cations on the electrode reaches a local maximum and then starts to decrease monotonically. Note that the competition between the Coulomb and dielectrophoretic forces acting on the cations does not violate the monotonic growth in the anion concentration on the electrode taking place within the model with nonpolarizable and nonpolar ions leading only to the inflection point on its voltage dependence. However, the anion concentration on the electrode increases much more slowly with voltage than the local concentration of the cations in the region of negative voltages. Additional dielectrophoretic attraction of the cations to the electrode makes it more difficult for the anions to expel the cations from the near-surface layer at positive potentials than in the case of replacement of the anions with cations at negative voltages. In the region of negative potential drops, with an increase in the absolute value, we observe monotonic growth in the cation concentration on the electrode up to the densest packing (saturation) of the cations in the near-surface layer, which occurs at less negative voltages with an increase in the cation dipole moment. Such behavior can be easily explained by the fact that an increase in the cation dipole moment results in a higher dielectrophoretic force that, in turn, enforces the attraction of the cations to the negatively charged electrode. We would also like to note that the value of the differential capacitance maximum shift to more negative voltages depends on the asymmetry in the local cation concentration on the electrode at negative voltages and anion concentration at positive voltages: the stronger the asymmetry of the concentrations, the bigger is the shift.
In the case of nonpolar and polarizable cations ($p_{+}=0$, $\alpha_{+}\neq 0$), we observe qualitatively different behavior of the ionic concentrations on the electrode. As is seen from Fig. \ref{fig:cpolIL}, at sufficiently small static polarizability of the cations, in the region of positive voltages, an increase in the potential drop results in the monotonic growth in the local anion concentration up to the full saturation of the near-surface layer and monotonic decrease in the cation concentration up to the total expulsion of the cations. However, when the static polarizability of the cations exceeds a certain threshold value, increase in the potential drop does not lead to the full expelling of the cations from the electrode surface anymore, and we observe a mixture of anions and cations in the fully saturated near-surface layer. Such surprising behavior is impossible in the ionic liquids with polar and nonpolarizable cations, because the latter possesses a fixed dipole moment value, in contrast to the polarizable cations, whose dipole moments increase linearly with the local electric field. Thus, a fine equilibrium between the dielectrophoretic attraction of the cations to the electrode, their Coulomb repulsion from the electrode, and the steric repulsive interactions of anions and cations in the near surface layer results in the saturation of the cation concentration on the positively charged electrode. 

\begin{figure}
\includegraphics[width=0.6\linewidth]{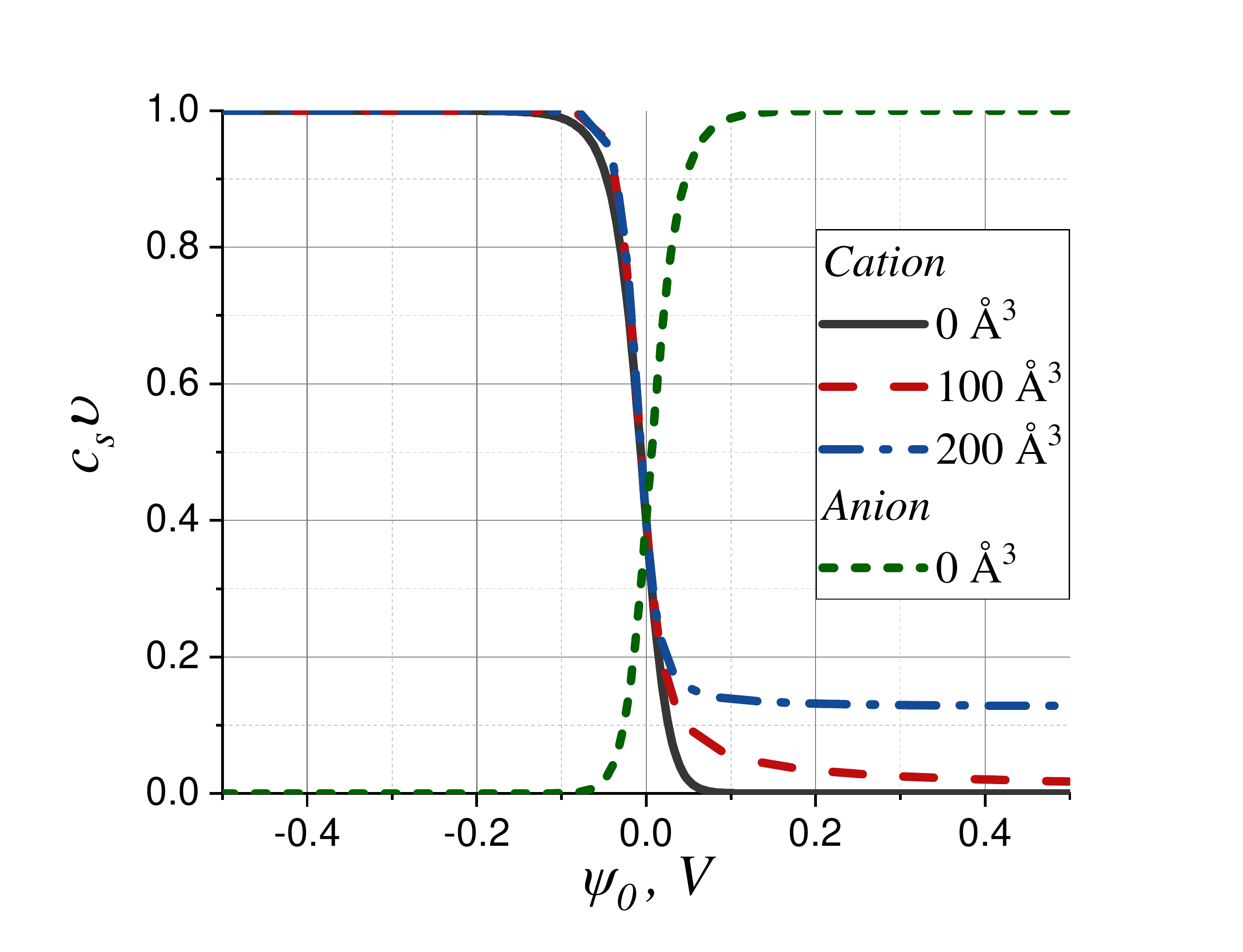} \\
\caption{The volume fraction of ions at different values of static polarizability $\alpha_{+}$ on the electrode surface. The data are shown for $\alpha_{-}=p_{+}=p_{-}=0$, $\varepsilon=5$, $v^{1/3}=5~{\angstrom}$, $cv =0.4$, $T=298~K$.}
\label{fig:cpolIL}
\end{figure}

\begin{figure}
\begin{tabular}{c c}
\includegraphics[width=0.5\linewidth]{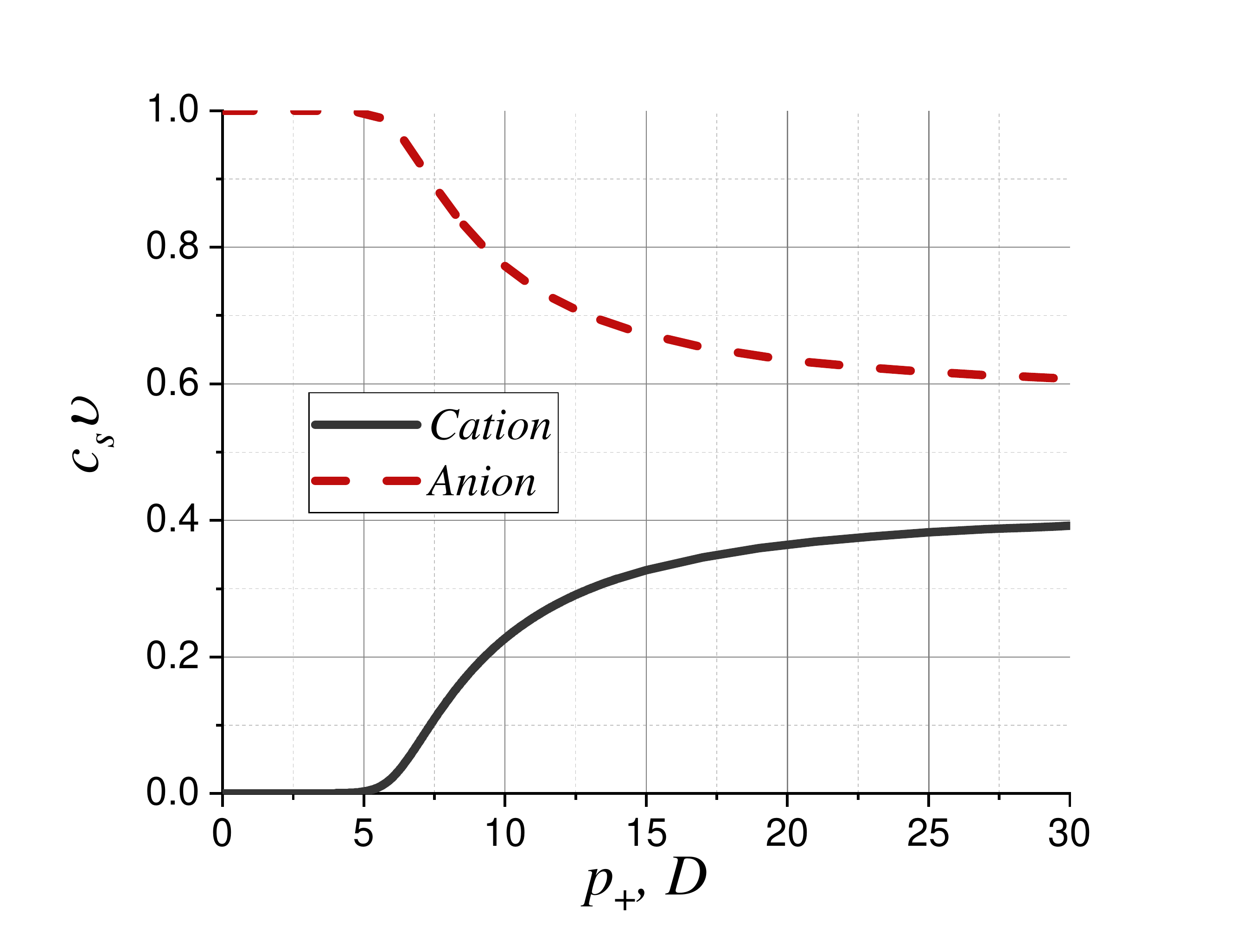}&\includegraphics[width=0.5\linewidth]{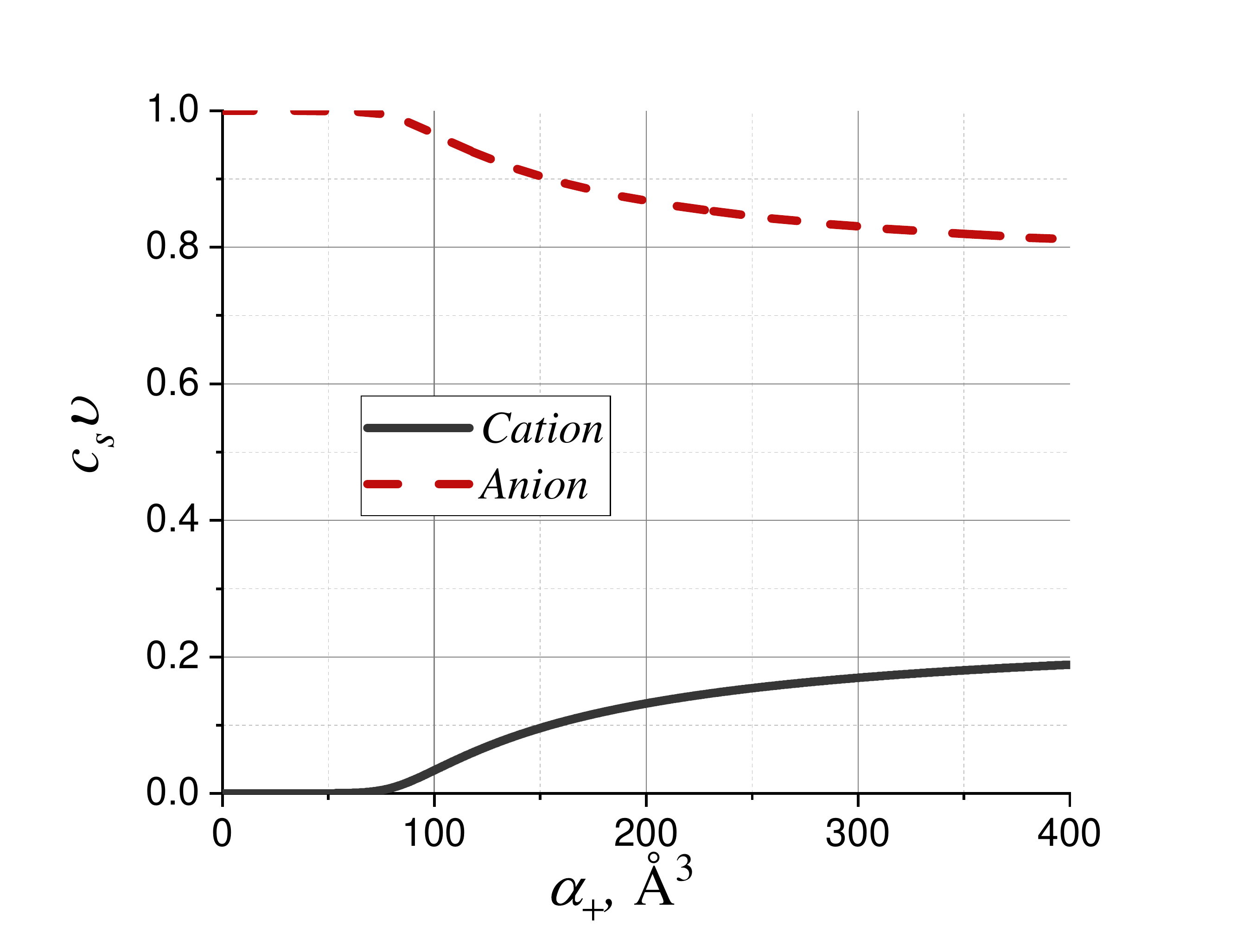} \\ [-0.3cm]
(a)&(b)\\
\end{tabular}
\caption{The volume fraction of the cations and anions on the electrode surface as a function of the cation dipole moment $p_{+}$ (a) and the static polarizability $\alpha_{+}$ (b) at a fixed potential $\psi_{0}=0.2~V$. The data are shown for $\alpha_{-}=\alpha_{+}=p_{-}=0$ (a), $\alpha_{-}=p_{+}=p_{-}=0$ (b), $\varepsilon=5$, $v^{1/3}=5~{\angstrom}$, $cv = 0.4$, $T=298~K$.}
\label{fig:Conc02V}
\end{figure}

Now, we would like to discuss the dependences of the ionic concentrations on the electrode on the dipole moment (Fig. \ref{fig:Conc02V}a) and the static polarizability (Fig. \ref{fig:Conc02V}b) of the cations at fixed positive applied voltage ($\psi_0=0.2~V$). As is seen from Fig. \ref{fig:Conc02V}, at sufficiently small values of both variables, at a rather big potential drop, there are absolutely no cations on the electrode surface, while the local concentration of the anions reaches the maximal value. However, when the dipole moment and static polarizability of cations exceed certain threshold values, the concentration of the cations (anions) starts to increase (decrease) achieving the plateau.    

\begin{figure}[!h]
\begin{tabular}{c c}
 \includegraphics[width=0.5\linewidth]{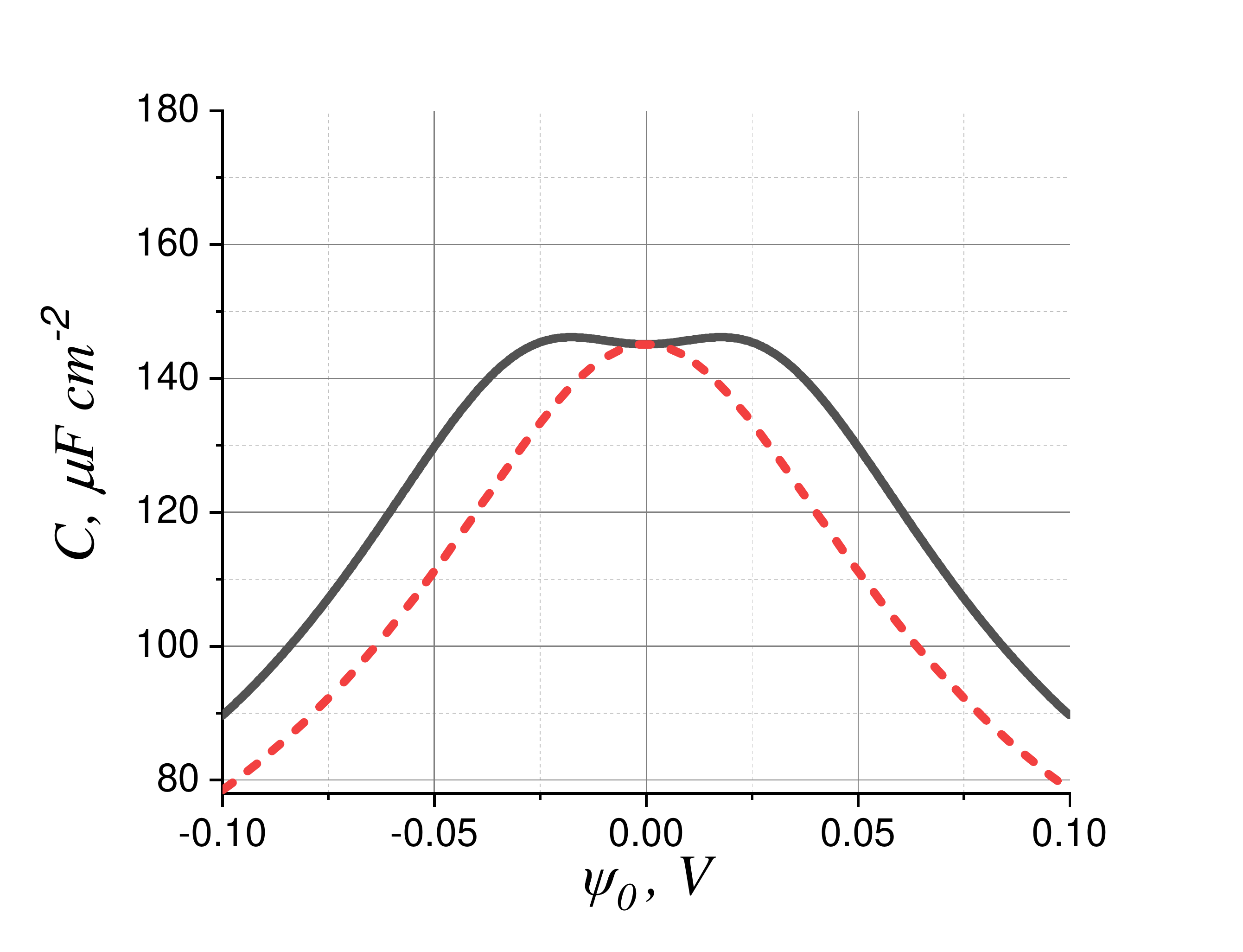}&\includegraphics[width=0.5\linewidth]{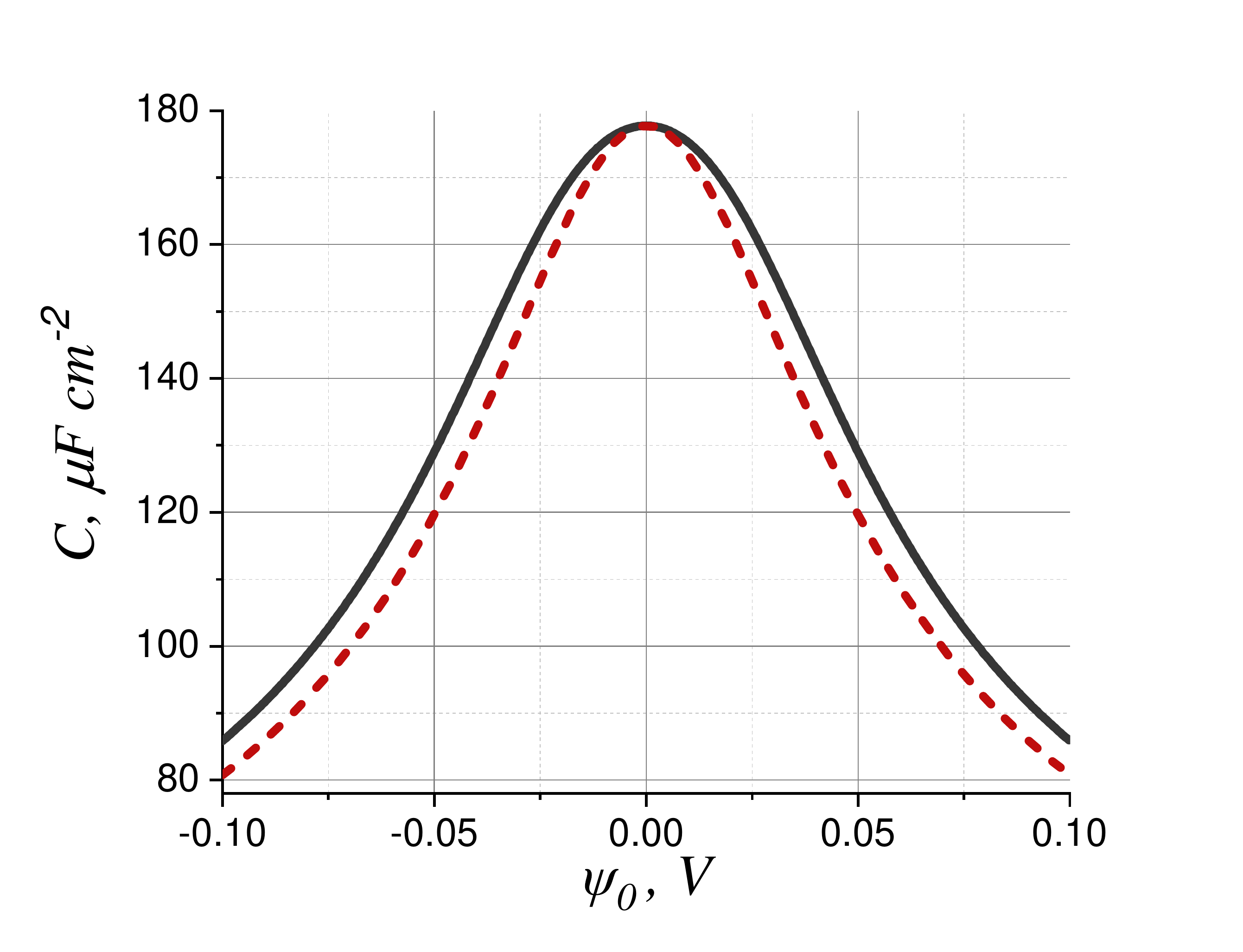} \\ [-0.3cm]
 (a)&(b)\\
\end{tabular}
\caption{Differential capacitance profiles calculated within the analytical theory taking into account the nonlinear polarization response of ions (full black line) and the theory accounting for the dielectric effect via the constant dielectric permittivity (dashed red line) for two bulk ionic concentrations. The data are shown for $cv = 0.3$ (a) and $cv = 0.4$ (b), $p_{+}=p_{-}=0$, $\alpha_{-}=\alpha_{+}=50~\angstrom^3$, $v^{1/3}=5~{\angstrom}$, $T=298~K$.}
\label{fig:nonlin}
\end{figure}

As it was mentioned in the introductory section, Lauw et al. \cite{lauw2009room} showed within the numerical polymer self-consistent field theory that inclusion of the nonlinear polarization effect originating from the dielectric mismatch of different ionic moieties in the ionic liquid model results in the camel-shaped curve of the differential capacitance instead of the expected bell-shaped profile predicted within the mean-field model with constant dielectric permittivity of the medium \cite{kornyshev2007double}. In this regard, it is interesting to investigate how the simultaneous nonlinear polarization effect of both ionic species affects the differential capacitance in the present model. For this purpose, we would like to consider the situation, when the cations and anions have equal electronic polarizabilities ($\alpha_{\pm}=\alpha$) and zero permanent dipole moments. Fig.~\ref{fig:nonlin}a demonstrates a comparison of the differential capacitance profiles, calculated within our analytical theory taking into account the nonlinear polarization response of ions and the theory accounting for the dielectric effect via constant dielectric permittivity determined by the expression $\varepsilon_{IL}=\varepsilon+2\alpha c/\varepsilon_{0}\approx 9$ (which can be obtained within the linear dielectric response approximation of the present theory) at $\alpha=50~\angstrom^3$ and $cv=0.3$. As is seen, in the first case, the present theory shows a symmetric camel-shaped profile as well as in Ref. \cite{lauw2009room}, while in the second case -- a bell-shaped one in accordance with the mean-field theory presented in Ref. \cite{kornyshev2007double}. However, at higher bulk ionic concentrations ($cv=0.4$, see Fig.~\ref{fig:nonlin}b), the nonlinear dielectric ion response does not violate the bell-shaped dielectric capacitance profile, only making the peak wider. We can conclude that the theory taking explicitly into account static polarizability of both ionic species predicts the transition from the camel-shaped profile to the bell-shaped one at higher bulk ionic concentrations than those predicted in the theory with nonpolarizable ions. We would like to mention that a similar picture is observed within the theory of polar ions ($p_{\pm}=p$, $\alpha_{\pm}=0$). 

It is instructive to discuss the asymptotic behavior of the differential capacitance at large absolute values of the potential drop (at $\beta q|\psi_0|\gg 1$) for the case of an ionic liquid with polarizable and polar ions following from the present theory. Using eqs. (\ref{loc_eq}) and (\ref{cap}), we obtain $C\simeq \sqrt{\pm q\varepsilon_{\mp}\varepsilon_0/2v\psi_0}$, where $\varepsilon_{\mp}\simeq\varepsilon+\alpha_{\mp}/\varepsilon_0 v$ is the local dielectric permittivity of the ionic liquid on the electrode at the densest packing of the ions in the near-surface layer; $\pm$ signs correspond to the positive and negative potential drops, respectively. As is seen, the dipole moments, $p_{\pm}$, drop out from the final expression due to the dielectric saturation effect taking place at rather large local electric fields. That is why the differential capacitance increases with polarizability (Fig. \ref{fig:CapIL}a) and practically does not change with the dipole moment increase (Fig.\ref{fig:CapIL}b) in the region of sufficiently high negative voltages. Thus, the mean-field theory accounting for the static polarizabilities and permanent dipole moments of the ionic species predicts qualitatively the same asymptotic dependence of the differential capacitance on the potential drop at $\beta q|\psi_0|\gg 1$ as in Kornyshev's mean-field theory \cite{kornyshev2007double} and classical DFT for the hard sphere mixture reference system \cite{jiang2011density}. Note that the aforementioned numerical polymer self-consistent field theory \cite{lauw2009room}, as it was reported by the authors, predicts a faster decay of the differential capacitance, namely, $C\sim |\psi_0|^{-0.6}$ for the bell-shaped profile and $C\sim |\psi_0|^{-0.8}$ for the camel-shaped one. Such asymptotic behavior may be explained by the special dendrimeric model of the ionic species adopted by the authors.

\subsection{Electrolyte solution case}
Now, we would like to discuss the differential capacitance behavior and the ionic concentrations on the electrode predicted by the present model in the case of sufficiently dilute electrolyte solutions ($cv\approx 0.01$, $v^{1/3}=5~\angstrom$) with polarizable or polar cations. This situation might be realized for an ionic liquid dissolved in some organic solvent like acetonitrile with certain dielectric permittivity, which is assumed to be $\varepsilon\approx 40$. Thus, we consider the solvent as a continuous dielectric medium, neglecting the crowding, polarizability, and permanent dipole moment of the polar solvent molecules. These effects were taken into account in paper \cite{iglivc2010excluded}.
\begin{figure}
\begin{tabular}{c c}
 \includegraphics[width=0.5\linewidth]{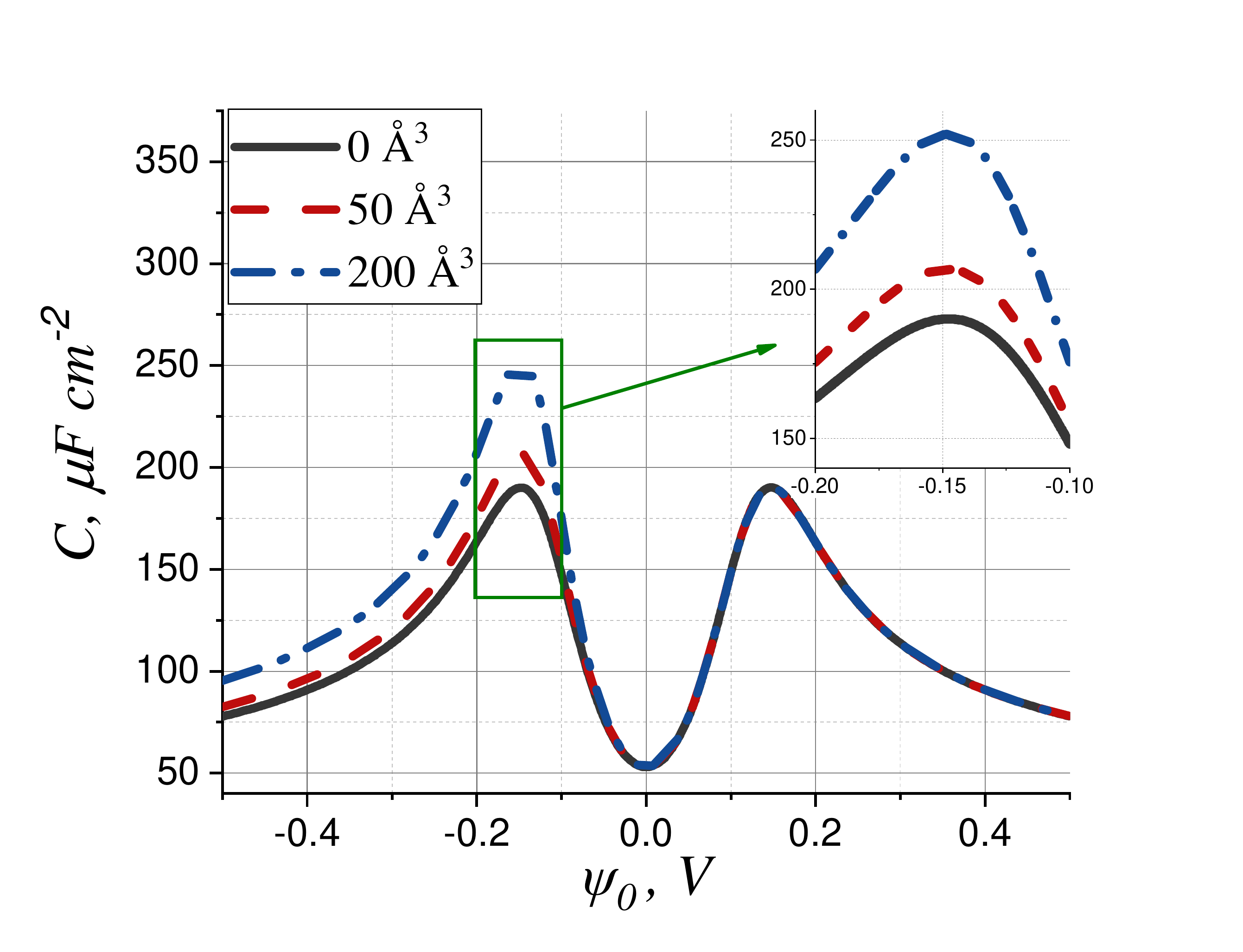}&\includegraphics[width=0.5\linewidth]{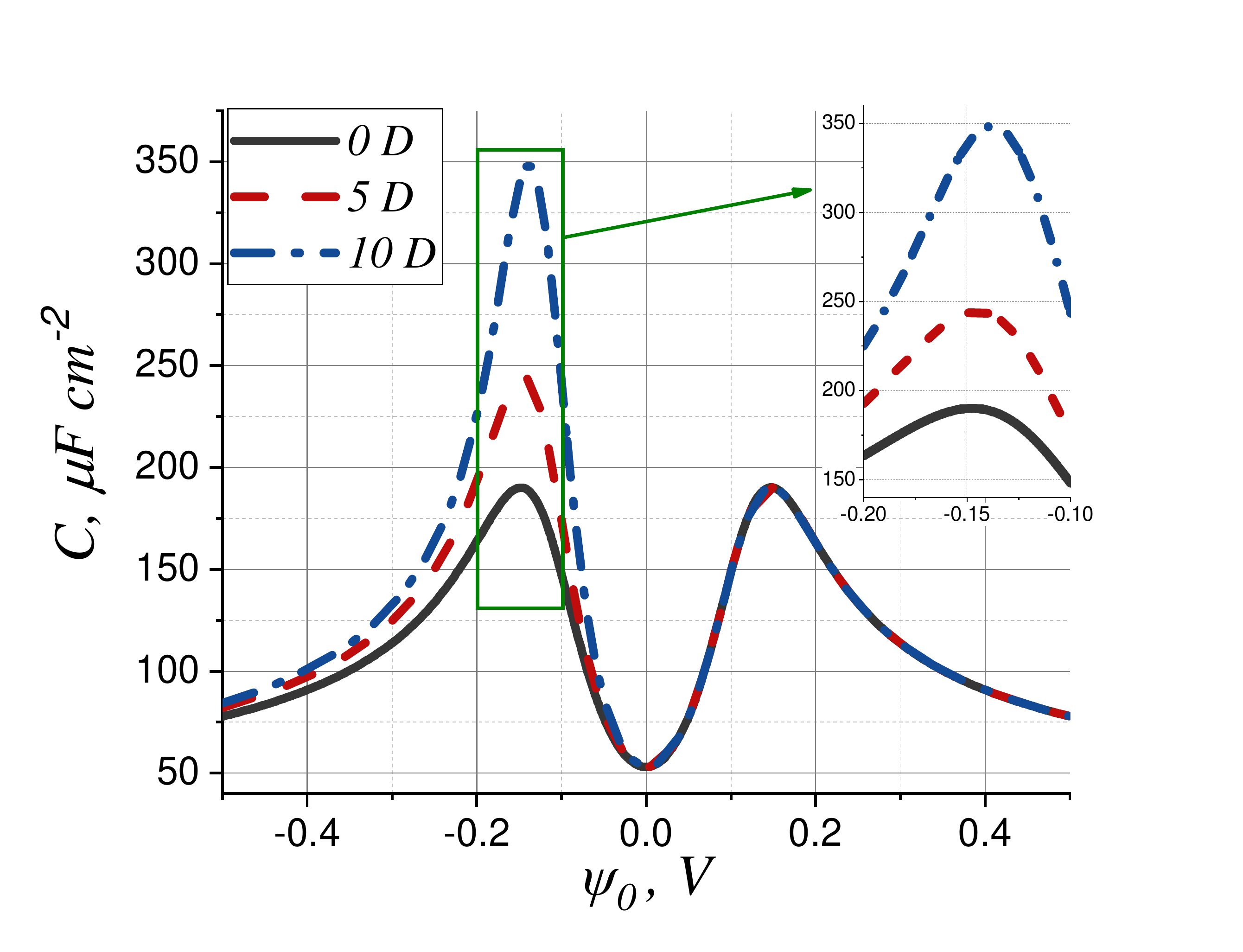} \\ [-0.3cm]
 (a)&(b)\\
\end{tabular}
\caption{Differential capacitance of the EDL in an electrolyte solution as a potential drop function at different values of electronic polarizability $\alpha_{+}$ (a) and dipole moment $p_{+}$ (b). The data are shown for $\alpha_{-}=p_{+}=p_{-}=0$ (a), $\alpha_{-}=\alpha_{+}=p_{-}=0$ (b), $cv \approx 0.01$, $\varepsilon=40$, $v^{1/3}=5~{\angstrom}$, $T=298~K$.}
\label{fig:CapES}
\end{figure}

\begin{figure}
\includegraphics[width=0.6\linewidth]{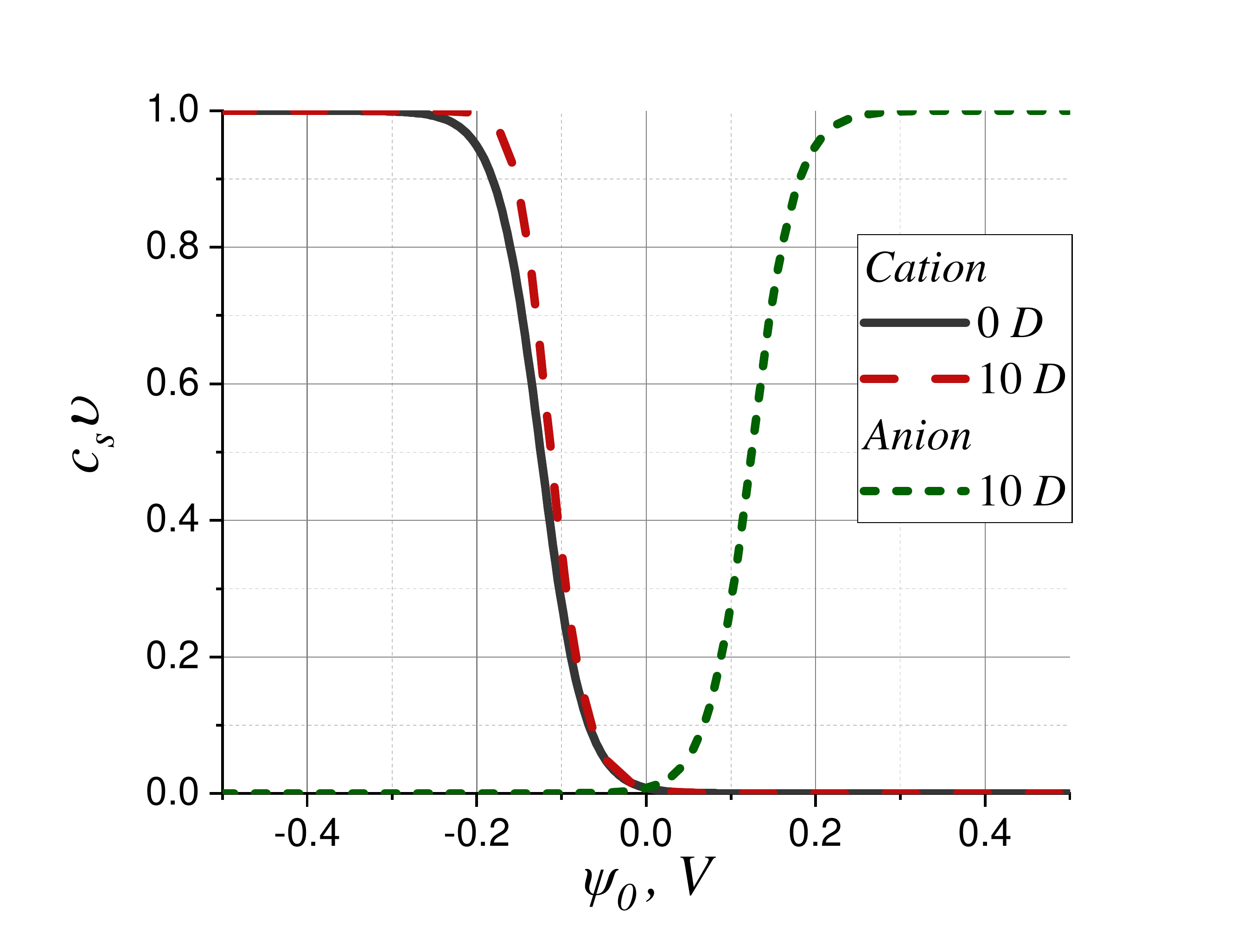} \\
\caption{Volume fractions of the ions at different dipole moment values, $p_{+}$, on the electrode surface in the electrolyte solution. The data are shown for $\alpha_{-}=\alpha_{+}=p_{-}=0$, $cv = 0.008 $, $\varepsilon=40$, $v^{1/3}=5~{\angstrom}$, $T=298~K$.}
\label{fig:ConcES}
\end{figure}

Figures \ref{fig:CapES}a,b illustrate the differential capacitance dependences on the potential drop for an electrolyte solution with polarizable (a) and polar (b) cations. We can see that the differential capacitance profiles have a camel shape with asymmetric peaks. As is seen, an increase in the polarizability or dipole moment of the cations leads to monotonic growth in the differential capacitance in the region of negative voltages and shifts the maximum position to less negative values of the surface potentials. The small shift of the differential capacitance maximum is explained by the fact that higher polarizability or dipole moment of the cations lead to EDL saturation at lower and lower absolute values of the potential drop (Fig. \ref{fig:ConcES}). As is seen, in contrast to the ionic liquid case, in the case of an electrolyte solution, we observe rather trivial behavior of the local ionic concentrations on the electrode -- their monotonic increase with the absolute value of voltage up to the saturation of the near-surface layer.

\section{Conclusions}
In conclusion, we have developed a general mean-field theory of a flat diffuse electric double layer in ionic liquids and electrolyte solutions with polarizable and polar ions on a charged electrode. In the framework of the formulated theory, we have for the first time derived a general analytical expression for electric double layer differential capacitance determining it as the absolute value of the ratio of the ionic local charge density to the local electric field on the electrode. The local electric field as a potential drop function is obtained from the ionic liquid mechanical equilibrium condition on the electrode, while the local charge density of the ions -- via the mean-field expressions for the local ionic concentrations as functions of voltage and local electric field. Based on the obtained analytical expression, we analyzed the behavior of differential capacitance as a function of voltage with an increase in the static polarizability and permanent dipole moment of the cations for ionic liquid and rather dilute electrolyte solution cases. In the ionic liquid case, we determined that an increase in polarizability or permanent dipole moment of the cations makes the differential capacitance at negative surface potentials higher and shifts its maximum to more negative voltages. We also found that for the competition between the dielectrophoretic attraction and the Coulomb repulsion acting on the polar cations results in nonmonotonic behavior of the cation concentration on the electrode with a voltage increase. At the same time, an increase in the electronic polarizability or permanent dipole moment of the cations does not disrupt the monotonic increase in the anion concentration on the electrode with voltage, which is observed in the model with nonpolarizable and nonpolar ions \cite{kornyshev2007double}, but leads only to the inflection point on its voltage dependence in case of polar cations. We have shown that in the case, when both ions are polarizable or polar ones, the camel-shaped differential capacitance profile transforms into a bell-shaped profile at higher bulk ionic concentrations than those predicted by the theory with nonpolarizable and nonpolar ions immersed in a homogeneous dielectric medium.

For the electrolyte solution case, we have established that an increase in the electronic polarizability or dipole moment of the cations leads to a more asymmetric camel-shaped capacitance profile. We have also observed that the higher maximum (in the region $\psi_{0}<0$) not only grows, but also slightly shifts to less negative voltages with an increase in the polarizability or dipole moment of the cations. We showed that the latter is related to the fact that the increase in the dipole moment or electronic polarizability of the cations makes their attraction to the negatively charged electrode stronger, which, in turn, leads to the saturation of near-surface layer at a smaller absolute voltage value. We have found that, in contrast to the ionic liquid case, in the case of an electrolyte solution, the increase in the static polarizability or dipole moment of the cations does not violate the monotonic behavior of the ionic concentrations with a voltage increase.

We believe that the theoretical model formulated in the present paper and the obtained theoretical findings could be useful in numerous electrochemical applications, such as batteries, supercapacitors, electrodeposition, catalysis, {\sl etc}.

\section*{Acknowledgments}
The research is partially supported by the grant of the President of the Russian Federation (project No. MD-341.2021.1.3). The theory presented in section 2 is supported by the Russian Science Foundation (Grant No. 21-11-00031). The authors thank anonymous Referees for valuable comments and suggestions allowed to improve the text.
\bibliography{name}
\end{document}